\documentclass[12pt]{article}

\usepackage[utf8]{inputenc}
\usepackage[english]{babel}
\usepackage{amsmath}
\usepackage{amsfonts}
\usepackage{amssymb}
\usepackage{amsthm}
\usepackage{colonequals}
\usepackage{multicol}
\usepackage{nccmath}
\usepackage{makecell}
\usepackage{braket}
\usepackage{bussproofs}

\usepackage[capitalise]{cleveref}

\theoremstyle{plain}
\newtheorem{theorem}{Theorem}
\newtheorem{lemma}[theorem]{Lemma}

\newtheorem{remark}[theorem]{Remark}

\theoremstyle{definition}
\newtheorem{definition}[theorem]{Definition}
\newtheorem{example}[theorem]{Example}

\newcommand{\justifies}{:}
\newcommand{\model}{\mathcal{M}} 
 
\newcommand{\valuation}{\varepsilon}

\newcommand{\force}{\Vdash}
\newcommand{\prove}{\vdash}
\newcommand{\notforce}{\nVdash}
\newcommand{\notprove}{\nvdash}

\newcommand{\CS}{\mathsf{CS}}

\newcommand{\je}{\mathsf{JE}}
\newcommand{\jem}{\mathsf{JEM}}

\newcommand{\logicje}{\je_{\CS}}
\newcommand{\logicjem}{\jem_{\CS}}

\newcommand{\maximal}{\mathsf{M}_\je}

\newcommand{\MP}{\bf{MP}}
\newcommand{\axnec}{\bf{AN}_\CS}

\newcommand{\jplus}{\textbf{j+}_1}
\newcommand{\djplus}{\textbf{j+}_2}
\newcommand{\jt}{\textbf{jt}}

\newcommand{\pset}{\mathcal{P}}
\newcommand{\pterms}{\mathsf{PTm}}
\newcommand{\jterms}{\mathsf{JTm}}
\newcommand{\terms}{\mathsf{Tm}}
\newcommand{\propositions}{\mathsf{Prop}}
\newcommand{\formulas}{\mathsf{Fm}}

\newcommand{\modalE}{\mathsf{E}}
\newcommand{\modalM}{\mathsf{EM}}
\newcommand{\modalL}{\mathsf{M}}
\newcommand{\justificationL}{\mathsf{J}}

\newcommand{\e}{\mathsf{e}}
\newcommand{\m}{\mathsf{m}}

\makeatletter
\newcommand{\todo}[1]{\marginpar{\textbf{TODO\footnotemark}}\@latex@warning{TODO: #1}\footnotetext{ #1}}

\makeatother

\author{Atefeh Rohani, Thomas Studer\\ \small{Institute of Computer Science, University of Bern, Switzerland}}

\title{Explicit non-normal modal logic\thanks{This work was supported by the Swiss National Science Foundation grant 200020$\_$184625.}}
\date{}

\begin{document}

\maketitle

\begin{abstract}
Faroldi argues that deontic modals are hyperintensional and thus traditional modal logic cannot provide an appropriate formalization of deontic situations.  To overcome this issue, we introduce novel justification logics as hyperintensional analogues to non-normal modal logics. We establish soundness and completeness with respect to various models and we study the problem of realization.
\end{abstract}

\section{Introduction}

Justification logic~\cite{artemovFittingBook,justificationLogic2019} is a variant of modal logic that replaces the implicit $\Box$-operator with 
explicit justifications.  Instead of formulas $\Box A$, meaning, e.g.,  \emph{$A$ is known} or \emph{$A$ is obligatory}, the language of justification logic features formulas of the form $t:A$ that stand for \emph{$t$ justifies the agent's knowledge of~$A$} or  \emph{$A$ is obligatory for reason $t$}, where $t$ is a so-called justification term.

The first justification logic, the Logic of Proofs~\cite{Art01BSL}, has been developed by Artemov in order to provide a classical provability semantics for the modal logic $\mathsf{S4}$ (and thus also for intuitionistic logic)~\cite{Art01BSL,weakArithm}. Starting with the work of Fitting~\cite{Fit05APAL}, several interpretations of justification logic have been presented that combine justifications with traditional possible world models~\cite{Art12SLnonote,KuzStu12AiML,StuderLehmannSubsetModel2019}. This opened the door for numerous applications of justification logic, e.g., in epistemic and deontic contexts~\cite{Art06TCS,BucKuzStu11JANCL,komaogst,Ren08PhD,Stu13JSL,KnowingWhy}.

One of the features of a normal modal logic is that it is closed under the rule of necessitation, that is if $F$ is valid, then so is $\Box F$.  Hence together with axiom $\mathsf{K}$, we can easily derive the rule of monotonicity:
Suppose $A \to B$ is valid.  By necessitation, we get $\Box (A\to B)$. By axiom $\mathsf{K}$ and modus ponens we conclude $\Box A \to \Box B$. 

Pacuit~\cite{Pacuit2017} mentions several interpretations of $\Box$ for which the validities and rules of inference of normal modal logic can be questioned. A well-known example is the paradox of gentle murder~\cite{Forrester1982}, where $\Box$ is read as \emph{ought to}. 
Consider the statements:
\begin{align}
&\text{Jones murders Smith.} \label{eq:gm:1}\\
&\text{Jones ought not to murder Smith.} \label{eq:gm:2}\\
&\text{If Jones murders Smith, then Jones ought to murder Smith gently.} \label{eq:gm:3}
\end{align}
These sentences seem to be consistent. However, from \eqref{eq:gm:1} and \eqref{eq:gm:3} we infer 
\begin{equation}
\text{Jones ought to murder Smith gently.} \label{eq:gm:4}
\end{equation}
Moreover, we have the following valid implication
\begin{equation}
\text{If Jones murders Smith  gently, then Jones  murders Smith.} \label{eq:gm:5}
\end{equation}
By the rule of monotonicity, \eqref{eq:gm:5} implies
\begin{multline}
\text{If Jones ought to murder Smith gently,}\\ 
\text{then Jones  ought to murder Smith.} \label{eq:gm:6}
\end{multline}
Now \eqref{eq:gm:4} and  \eqref{eq:gm:6} together yield
\begin{equation}
\text{Jones  ought to murder Smith.} \label{eq:gm:7}
\end{equation}
This contradicts  \eqref{eq:gm:2}. This argument suggests that deontic modal logic should not validate the rules of normal modal logic and thus a semantics different  from Kripke semantics is needed. The  traditional approach for models of non-normal modal logics is to use neighborhood semantics. There, a so-called neighborhood function $N$ assigns to each world $w$ a set of sets of worlds $N(w)$ and a formula $\Box F$ is true at $w$ if the truth set of $F$ is an element of $N(w)$.

Justification logics are parametrized by a constant specification, which is a set 
\begin{align*}
\CS \subseteq \{ (c,A) \ |\ &\text{$c$ is a constant justification term and}\\ 
&\text{$A$ is an axiom of justification logic}\}  .
\end{align*}
A constant specification $\CS$ is called {\it axiomatically appropriate} if for each axiom $A$ there is a constant $c$ such that $(c, A) \in \CS$. 
Instead of the rule of necessitation, justification logics include a rule called \emph{axiom necessitation} saying that one is allowed to infer $c\justifies A$ if $(c,A) \in \CS$.
Hence, In epistemic settings, we can calibrate the reasoning power of the agents by adapting the constant specification.
\footnote{It is important to consider axiom necessitation as a rule and not an axiom schema, even though it is a rule without premises. If we considered $c \justifies A$ as an axiom for each $(c,A) \in \CS$, then the notion of an axiom would depend on the notion of a constant specification, which depends on the notion of an axiom. In order to avoid this circularity, we introduce axiom necessitation as a rule. }

Faroldi and Protopopescu~\cite{FaroldiBook2019,FaroldiHyperPracticalReasons} suggest using this mechanism also in deontic settings in order to avoid the usual paradoxes. For instance, they discuss Ross' paradox~\cite{Ross41}, which is:
\begin{equation}\label{eq:ross:1}
\text{You ought to mail the letter.}
\end{equation}
implies
\begin{equation}\label{eq:ross:2}
\text{You ought to mail the letter or burn it.}
\end{equation}
The reason is as before. It is a classical validity that
\begin{equation}\label{eq:ross:3}
\text{\emph{you mail the letter} implies \emph{you mail the letter or burn it}.}
\end{equation}
By the monotonicity rule we find that \eqref{eq:ross:1} implies \eqref{eq:ross:2}.

Fardoli and Protopopescu avoid this paradox by restricting the constant specification such that although \eqref{eq:ross:3} is a logical validity, there will no justification term for it. Thus the rule of monotonicity cannot be derived and there is no paradox.

One of the reasons why Faroldi prefers justification logic over using neighborhood  models is that he claims that deontic modalities are \emph{hyperintensional}~\cite{FaroldiDeonticModals}, i.e.~they can distinguish between logically equivalent formulas. Here is an example to show that the usual modal operator is not hyperintensional. Consider the following sentences:
\begin{equation}\label{eq:hyp1}
\text{You ought to drive. }
\end{equation}
\begin{equation}\label{eq:hyp2}
\text{You ought to drive or to drive and drink.}
\end{equation}
Intuitively sentences \eqref{eq:hyp1} and \eqref{eq:hyp2} are not equivalent, yet their formalizations in  modal logic are so. If we represent \eqref{eq:hyp1} by $\Box A$ and~\eqref{eq:hyp2} by $\Box (A \vee (A \wedge B))$,  then we have $A \leftrightarrow A \vee (A \wedge B)$ by propositional reasoning and by the rule of equivalence we infer $\Box A \leftrightarrow \Box (A \vee (A \wedge B))$.
However, hyperintensionlity is one of the distinguishing features of justification logics: they are hyperintensional by design. Even if $A$ and $B$ are logically equivalent, we may have that a term $t$ justifying $A$ does not  justify $B$. 
Think of the Logic of Proofs, where the  terms represent proofs in a formal system (like Peano arithmetic). Let $A$ and $B$ be logically equivalent formulas. In general, a proof of $A$ will not also be a proof of $B$.  In order to obtain a proof of $B$ we have to extend the proof of $A$ with a proof of $A \to B$ and an application of modus ponens. 
Thus in  justification logic, terms do distinguish between equivalent formulas, which, according to Faroldi, makes it a suitable framework for deontic reasoning. 

There is a problem with restricting the constant specification. Namely, the resulting constant specification will not be \emph{axiomatically appropriate}, i.e.~there will be axioms that are not justified by any term. This implies, however, that the Internalization property (saying that a justification logic internalizes its own notion of proof) does not hold, which is a problem for several  reasons.

First, Internalization is needed to obtain completeness with respect to fully explanatory models.  That is models where each formula that is obligatory (or believed) in the sense of the modal $\Box$ operator has a justification.

Further, Internalization is often required to obtain completeness when a form of the $\mathsf{D}$ axiom is present~\cite{KuzStu12AiML,StuderLehmannSubsetModel2019,Pac05PLS}. 
In deontic settings, this is often the case since obligations are supposed not to contradict each other. 
Hence restricting the constant specification leads to deductive systems that are not complete.
Conflicting obligations in justification logic have been studied in~\cite{ConflictingObligations}. Recently, it turned out that this approach can also be used to analyze an epistemic paradox of quantum physics~\cite{StuderQuantum}.

Moreover, Internalization is essential to obtain realization results. A justification logic realizes a  modal logic if, given any theorem $F$ of the modal logic,  each occurrence of $\Box$ in $F$ can be replaced with some justification term such that the resulting formula is a theorem of the justification logic. Realization is an important property connecting implicit and explicit modalities.  

In the present paper, we introduce two novel justification logics $\logicje$ and $\logicjem$ that are the explicit counterparts of the non-normal modal logics $\modalE$ and $\modalM$, respectively. As usual for justification logics, $\logicje$ and $\logicjem$ are hyperintensional and can therefore serve as an appropriate formalization of deontic modals.
On a technical level, the main novelty of our paper is the introduction of two types of terms for $\logicje$ and $\logicjem$. 
This makes it possible to formalize the characteristic principle of  $\logicje$ and $\logicjem$ as an axiom (and not as a rule) and, therefore, our logics have the Internalization property.  Note  that we are not the first to use two types of terms. In~\cite{KuzMarStr17}, terms for representing proofs and terms justifying consistency have been combined in constructive justification logic.

We show soundness and completeness of $\logicje$ and $\logicjem$ with respect to basic models, modular models and fully explanatory modular models.
Moreover, we show that the justification logics $\logicje$ and $\logicjem$ realize the modal logics $\modalE$ and $\modalM$, respectively. 
From a technical perspective, the case of realizing $\modalE$ is particularly interesting because there we have to deal with a rule that does not respect the polarities of subformulas.

{\bf Acknowledgement.} We thank the anonymous reviewers for their helpful comments.

\section{Justification logic}
To define the language of  our novel justification logic  $\logicje$, we extend the usual language of justification logic by introducing two types of terms. 
We consider \emph{proof terms} and \emph{justification terms}, which are inductively built-up from countably many proof constants and variables.
So if we denote proof constants by $\alpha_i$ and proof variables by $\xi_i$, the set of proof terms is defined inductively as follows: 
\[
 \lambda \coloncolonequals \alpha_i \ | \ \xi_i \ | \  (\lambda \cdot \lambda) \ | \ (\lambda+\lambda)   \ | \ !\lambda \ .
\]
Justification terms have the following form:
\[
t \coloncolonequals  \e(\lambda)    \ .
\]
where  $\lambda$ is a proof term. We denote the set of proof terms by $\pterms$ and the set of justification terms by $\jterms$. Therefore, the set of all terms is $ \terms := \pterms \cup \jterms$. 
We use $\lambda, \kappa, \gamma$ for elements of $\pterms$ and $r, s, t$ for elements of $\jterms$.

Let $\propositions$ be a countable set of atomic propositions.
Formulas are inductively defined as follows:
\[
F \coloncolonequals P_i \ | \  \bot \ | \ (F \to F) \ |\ \lambda \justifies F \  | \  [t] F \ ,
\]
where $P_i \in \propositions , \ \lambda \in \pterms , \ \text{and} \ t \in \jterms$.
We use $\formulas$ for the set of formulas. $\lambda \justifies F$ is read as {\it$\lambda$ proofs $F$} and $[t]F$ is read as {\it $t$ justifies $F$}. 
The axioms of $\je$ are:
\begin{fleqn}
\begin{equation}
\begin{array}{ll}\nonumber
 {\bf j} &\lambda \justifies (F \to G) \to ( \kappa \justifies F \to \lambda \cdot \kappa \justifies G) \\
\jplus\phantom{m} &( \lambda \justifies F \vee \kappa \justifies F) \to (\lambda+\kappa) \justifies F \\
\jt& \lambda \justifies F \to F \\
 {\bf j4} &  \lambda \justifies F \to\  !\lambda \justifies \lambda \justifies F \\
{\bf je} &  (\lambda \justifies (F \to G) \wedge \lambda \justifies (G \to F)) \to ( [\e(\lambda )]F \to [\e(\lambda)]G )  \\
{\textbf {je+}} & ([\e(\lambda)]F \vee [\e(\kappa)]F) \to [\e(\lambda + \kappa)]F \\
\end{array}
\end{equation}
\end{fleqn}
Note that the axioms {\bf j}, $\jplus$, $\jt$, and  {\bf j4} are exactly the axioms of the Logic of Proofs.  Indeed, dropping ${\bf je}$ and ${\textbf {je+}}$ from $\logicje$ and restricting the language to proof terms (hence excluding justification terms) yields the Logic of Proofs. 

Axiom {\bf je} shows how justification terms $\e(\lambda )$ are constructed  based on proof terms $\lambda$; and axiom {\textbf {je+}} is similar to axiom $\jplus$ but for justification terms. It shows that the operation $+$ combines two proof terms such that if $\e(\lambda)$ or $\e(\kappa)$ provides evidence for $F$, the combined evidence $\e(\lambda + \kappa)$ remains evidence for $F$. 

As we will see later, the axiom {\textbf{je+}} is only used to prove completeness of the logic $\je$ w.r.t.~fully explanatory models.  It is not needed to establish our other (completeness) results.

In order to define the deductive system for our logic, we first need the notion of a constant specification.

\begin{definition}[Constant Specification] 
A \emph{constant specification} $\CS$ is any subset:
\[
\CS \subseteq \{ \alpha \justifies A  \ | \ \alpha \text{ is a proof constant and }  A  \text{ is an axiom of $\je$} \} \ .
\]
A constant specification $\CS$ is called \emph{axiomatically appropriate} if for each axiom~$A$ of  $\je$ there is a constant $\alpha$  with  $(\alpha, A) \in \CS$.
\end{definition}

\begin{definition}[Logic $\logicje$] 
For a constant specification $\CS$, the logic $\logicje$ is defined by a Hilbert-style system with the axioms $\je$ and the inference rules modus ponens \textnormal{(}{\MP}\textnormal{)} and axiom necessitation \textnormal{(}{$\axnec$}\textnormal{)}, given by:
\[
\frac{}{\alpha \justifies A} \ \text{ where} \ (\alpha, A) \in \CS \ . 
\] 
\end{definition}
We write $\logicje \prove A$ to express that a formula $A$ is provable in $\logicje$. If the deductive system is clear from the context and we only want to stress the constant specification, we simply use $\prove_{\CS} A$. When the constant specification does not matter or is clear from the context, we drop the subscript $\CS$ and write  $\prove A$.

It is a standard result that justification logics with an axiomatically appropriate constant specification internalize their own notion of proof~\cite{Art01BSL,artemovFittingBook,justificationLogic2019}.

\begin{lemma}[Internalization]\label{l:internalization:1} 
Let\/ $\CS$ be an axiomatically appropriate constant specification. For any formula $A$ with $\prove A$, there exists a proof term $\lambda$ such that $\prove \lambda: A$.
\end{lemma}

Moreover, justification logics enjoy a deduction theorem~\cite{Art01BSL,artemovFittingBook,justificationLogic2019}.
\begin{lemma}[Deduction]
Let\/ $\CS$ be an arbitrary constant specification. 
For any set $\Delta$ of  formulas and for any formulas $A$ and $B$,
\[
\Delta, A \vdash B \quad\text{if{f}}\quad \Delta \vdash A \to B \ .
\]
\end{lemma}

Let us now turn to semantics. In order to present basic evaluations for $\logicje$, we need some operations on sets of formulas.

\begin{definition}\label{d:semOPs:1}
Let $X,Y$ be sets of formulas and $\lambda$ be a proof term. We define the following operations:
\begin{description}
\item $\ \lambda \justifies X \colonequals \{ \lambda \justifies F \ | \ F \in X \} $;
\item $X \cdot Y \colonequals \{ F  \ | \ G \to F \in X  \text{ for some} \ G \in Y \} $;
\item $X \odot Y \colonequals \{  F  \ | \  F \to G \in X \text{ and } G \to F \in X \ \text{for some} \ G \in Y \} \ .$
\end{description}
\end{definition}

\begin{definition}[Basic evaluation] 
Let $\CS$ be an arbitrary constant specification. A \emph{basic evaluation} for $\logicje$ is a function $\valuation$ that maps atomic propositions to 0 or 1 
\[ \valuation (P_i) \in \{0,1\} \ \text{for} \ P_i \in \propositions  \]
and maps terms to a set of formulas: 
\[ \valuation \colon \pterms \cup \jterms \to  \pset (\formulas) \ , \] 
such that for arbitrary $\lambda , \kappa \in \pterms$:
\begin{enumerate}
\item $\valuation(\lambda) \cdot \valuation(\kappa) \subseteq \valuation(\lambda\cdot \kappa)$;
\item $\valuation(\lambda) \cup \valuation(\kappa) \subseteq \valuation(\lambda + \kappa)$;
\item $ F \in \valuation(\lambda) \ \text{if} \ (\lambda, F) \in \CS$;
\item $\lambda \justifies \valuation(\lambda) \subseteq \valuation(!\lambda)$;
\item $\valuation(\lambda) \odot \valuation(\e(\lambda)) \subseteq \valuation(\e(\lambda))$;
\item $\valuation(\e(\lambda)) \cup \valuation(\e(\kappa)) \subseteq \valuation(\e (\lambda + \kappa))$. 
\end{enumerate}
\end{definition}

\begin{definition}[Truth under a basic evaluation]
We define truth of a formula~$F$ under a basic evaluation $\valuation$ inductively as follows:
\begin{enumerate}
\item $\valuation \notforce \perp$;
\item $\valuation \force P \ \text{iff} \ \valuation(P)=1  \text{ for }  P \in  \propositions$;
\item $\valuation \force F \to G \ \text{iff} \ \valuation \notforce F \ \text{or} \ \valuation \force G$;
\item$\valuation \force \lambda \justifies F \ \text{iff} \ F \in \valuation(\lambda)$;
\item $\valuation \force  [t]F \ \text{iff} \ F \in \valuation(t)$.
\end{enumerate}
\end{definition}

\begin{definition}[Factive basic evaluation]
A basic evaluation  $\valuation$ is called {\it factive} if for any formula $\lambda \justifies F$ we have  $\valuation \force \lambda \justifies F$ implies $\valuation \force F$. 
\end{definition}

\begin{definition}[Basic model] 
Given an arbitrary $\CS$,  a \emph{basic model} for $\logicje$ is a basic evaluation that is {\it factive}.
\end{definition}

As expected, we have soundness and completeness with respect to basic models. The following theorem is established in Appendix~\ref{ap:A}.

\begin{theorem}[Soundness and completeness w.r.t.~basic models] 
Let $\CS$ be an arbitrary constant specification.
The logic $\logicje$ is sound and complete with respect to basic models. 
For any formula~$F$, 
\[
\logicje \vdash F  \quad\text{if{f}}\quad
\varepsilon \Vdash F  \text{ for all  basic models  $\varepsilon$ for $\logicje$ } \ .
\]
\end{theorem}

\section{Neighborhood semantics and modular \newline models}

The main purpose of modular models is to connect justification logic to traditional modal logic. 
To define modular models for $\logicje$, we start with a neighborhood model (like for the modal logic $\modalE$) and assign to each possible world a basic evaluation.
This, however, is not enough since these basic evaluations may have nothing to do with the neighborhood structure of the model.
Hence we introduce the following principle:
\begin{multline*}
\text{having a specific justification for $F$ must yield}\\ \text{$F$ is obligatory in the sense of the neighborhood structure.}
\end{multline*}
This principle was first introduced in epistemic contexts and is, therefore, called \emph{justification yields belief} (JYB).

\begin{definition}[Neighborhood function]
For  a non-empty set of worlds~$W$, a neighborhood function is any $N \colon W \to \pset \left(  \pset (W) \right) $. 
\end{definition}

\begin{definition}[Quasi-model]
A  quasi-model for $\logicje$ is a triple \[\model = \langle W, N, \varepsilon \rangle\] where $W$ is a non-empty set of worlds, $N$ is a neighborhood function and $\varepsilon$ is an evaluation function that maps each world to a basic evaluation $\varepsilon_w$.
\end{definition}

\begin{definition}[Truth in quasi-model]
Let  $\model = \langle W, N, \varepsilon \rangle$ be a quasi-model. \emph{Truth of a formula at a world $w$} in a  quasi-model is defined inductively as follows: 
\begin{enumerate}
\item $\model , w \notforce \perp$;
\item $\model , w \force P \ \text{iff} \ \varepsilon_w (P) = 1, \ \text{for} \ P \in \propositions$;
\item $\model , w \force F \to G \ \text{iff} \ \model , w \notforce F \ \text{or} \ \model , w \force G$;
\item $\model, w \force \lambda \justifies F \ \text{iff} \ F \in \varepsilon_w (\lambda)$;
\item $\model, w \force [t] F \ \text{iff} \ F \in \varepsilon_w (t)$.
\end{enumerate}
We will write $\model \force F$ if $\model,w\force F$ for all $w\in W$. 
\end{definition}

\begin{remark}
The neighborhood function plays no rule in the definition of truth in quasi-models. Hence
truth in quasi-models is local to a possible world. Let $\model = \langle W, N, \varepsilon \rangle$ be a quasi-model. For any $w \in W$ and any formula $F$,
\begin{equation}\label{eq:locality:1}
 \model , w \force F \quad\text{if{f}}\quad \varepsilon_w \force F \ .
 \end{equation}
 \end{remark}

\begin{definition}[Factive quasi-model]
A  quasi-model  $\model = \langle W, N, \varepsilon \rangle$ is \emph{factive} if for each world $w$, we have that for 
any formula~$\lambda: F$, 
\[
\model , w \force \lambda: F \quad\text{implies}\quad \model , w \force  F \ .
\]
\end{definition}

\begin{definition}[Truth set]
Let  $\model = \langle W, N, \varepsilon \rangle$ be a quasi-model. The \emph{truth set} of a formula $F$, denoted by $|F| ^\model$,  is the set of all worlds in which $F$ is true, i.e.,
\[
|F|^\model := \{ \ w \in W \ | \ \model , w \force F \ \} \ .
\]  
\end{definition}

Further, we define
\[
\Box_w \colonequals \{  F \  | \  |F|^\model \in N(w) \}\ \ .
\]
Looking back at neighborhood models for $\modalE$, it is easy to see that $F \in \Box_w$ means  (modulo the different language that we are using) that $\Box F$ holds at world $w$.
As a result, we can formulate the principle of justification yields belief as follows:
\begin{equation}\label{eq:JYB:1}\tag{JYB}
\text{for any} \ t \in \jterms \ \text{and} \ w \in W, \text{ we have that }\valuation_w (t) \subseteq \Box_w \ .
\end{equation}

\begin{definition}[Modular model]
A \emph{$\logicje$  modular model} 
is a quasi-model for  $\logicje$  
that is factive and  satisfies~\eqref{eq:JYB:1}.
\end{definition}

$\logicje$ is sound and complete with respect to modular models. A proof of the following theorem is given in Appendix~\ref{ap:B}.

\begin{theorem}[Soundness and completeness w.r.t.~modular models] 
Let $\CS$ be an arbitrary constant specification. 
For each formula $F$ we have
\[
\logicje \prove F \quad\text{if{f}}\quad  \model \force F \text{ for all\/ $\logicje$  modular models $ \model $.}
 \]
\end{theorem}

It is natural to ask whether every obligatory formula in a modular model is justified by a justification term. 

\begin{definition}[Fully explanatory modular model]
A  $\logicje$  modular model $\model = \langle W, N, \varepsilon \rangle$ is \emph{fully explanatory} if for any $w \in W$ and any formula $F$, 
\[
|F|^\model \in N(w) \quad\text{implies}\quad F \in \varepsilon_w(t) \ \text{for some} \ t \in \jterms \ .
\]
\end{definition}

The fully explanatory property can be seen as the converse of justification yields belief. In fully explanatory models we have that for each world $w$,
\[
\bigcup _{t \in \jterms} \varepsilon_w(t) = \Box_w \ .
\]

For any axiomatically appropriate constant specification $\CS$, we can show that  $\logicje$  is sound and complete with respect to fully explanatory  $\logicje$ modular models. In order to obtain this, we need monotonicity of the $\e$-operation with respect to $+$ as expressed in axiom   {\textbf{je+}}. 
The proof is presented in Appendix~\ref{ap:C}.

\begin{theorem}[Soundness and completeness for fully explanatory modular models]
Let $\CS$ be an axiomatically appropriate constant specification. $\logicje$ is sound and complete with respect to  fully explanatory  $\logicje$ modular models.
\end{theorem}

\section{Monotonic justification logic}

There are several applications for which the modal logic  $\modalE$ is  too weak and one considers the extension of  $\modalE$  with the axiom $\Box (A \land B) \to (\Box A \land \Box B)$ or, equivalently, with the rule 
\[
\frac{A \to B}{\Box A \to \Box B} \  \ . 
\] 
The resulting logic is called $\modalM$. In this section we introduce an explicit counterpart $\jem$ of the modal logic $\modalM$.

First, we adapt the language as follows.  
Proof terms are given as before but without $+$:
\[
 \lambda \coloncolonequals \alpha_i \ | \ \xi_i \ | \  (\lambda \cdot \lambda)  \ | \ !\lambda \ .
\]
The set of \emph{justification terms} is built up inductively, starting from a countable set of justification variables $x_i$, by:
\[ t \coloncolonequals x_i   \ | \ t+t \ | \ \m(\lambda, t) 
\]
where $\lambda$ is a proof term. 
Formulas are then built using this extended set of justification terms. 
It will always be clear from the context whether we work with the basic language for $\je$ or with the extended language for $\jem$.

The axioms of $\jem$ consist of the axioms   {\bf j},  $\jt$, and  {\bf j4} together with
\begin{description}
\item {\bf jm} \hspace{5mm} \ \ $\lambda \justifies (F \to G) \to ([t]F \to [\m(\lambda, t)]G)$.
\item $\djplus$  \hspace{5mm} \ \ $ ([t] F \vee [s]F) \to [t+s]F$.
\end{description}
For a constant specification $\CS$, we now consider axioms of  $\jem$; and the system
$\logicjem$ consists of the axioms of $\jem$ plus the rules of modus ponens and axiom necessitation.
Note that Internalization and the Deduction theorem hold for $\logicjem$, too.
Axiom $\djplus$ will be used in the realization proof,  but we do not need $+$ for proof terms in $\jem$ and thus we dispense with axiom $\jplus$.
For $\jem$, we can establish completeness w.r.t.~fully explanatory models without using axiom {\textbf {je+}}, thus we do not include it in  $\jem$.

A \emph{basic evaluation for $\logicjem$} is defined similar to a basic evaluation for $\logicje$ with the conditions for $+$ on proof terms and for $\e$ dropped and with  the additional  requirements that
for arbitrary terms $\lambda \in \pterms$ and $t , s \in \jterms$:
\begin{enumerate}
\item $\valuation(\lambda) \cdot \valuation(t) \subseteq \valuation(\m(\lambda,t)) \ $;
\item $\valuation(t) \cup \valuation(s) \subseteq \valuation(t + s)$.
\end{enumerate}
Further we  define a \emph{monotonic basic model} (for $\logicjem$) as a basic evaluation for $\logicjem$ that is factive.

Similar to $\logicje$, we can show that $\logicjem$ is sound and complete with respect to monotonic basic models.

\begin{theorem}
Let $\CS$ be an arbitrary constant specification. The logic $\logicjem$ is sound and complete with respect to monotonic basic models. For any formula~$F$,
\[
\logicjem \vdash F \quad\text{if{f}}\quad 
\varepsilon \Vdash F  \text{ for all  monotonic basic models $\varepsilon$ for $\logicjem$ } \ .
\]
\end{theorem}

Now we are going to adapt modular models to $\logicjem$.
A neighborhood function $N$ for a non-empty set of worlds $W$ is called \emph{monotonic} provided that for each $w\in W$ and for each $X\subseteq W$, 
\[
\text{if $X \in N(w)$ and $X\subseteq Y \subseteq W$ then $Y \in N(w)$.}
\]
 
 A \emph{monotonic quasi-model} for $\logicjem$ is defined like a quasi-model for $\logicje$ but we use a monotonic neighborhood function and each world is mapped to a basic evaluation for  $\logicjem$.  
A \emph{monotonic modular model} is then defined like a modular model but the underlying quasi-model is required to be monotonic. As for $\logicje$ we get completeness or $\logicjem$ with respect to monotonic modular models.

\begin{theorem}
Let $\CS$ be an arbitrary constant specification. 
For each formula~$F$ we have
\[
\logicjem \prove F \quad\text{if{f}}\quad  \model \force F \text{ for all\/ $\logicjem$ monotonic modular models $ \model $.}
 \]
\end{theorem}

To achieve completeness with respect to fully explanatory monotonic modular models, one needs some additional construction to guarantee that the neighborhood function constructed in the canonical model is monotonic. 
Details can be found in Appendix~\ref{ap:Cmon}.

\begin{theorem}
 Let $\CS$ be an axiomatically appropriate constant specification. $\logicjem$ is sound and complete with respect to  fully explanatory  $\logicjem$ monotonic modular models.
\end{theorem}

\section{Realization}

This section is concerned with the exact relationship between some non-normal modal logic $\modalL$ and its explicit counterpart $\justificationL$.
Let $\formulas^\modalL$ denote the set of formulas from modal logic and  $\formulas^\justificationL$ the set of all justification logic formulas (for $\modalE$ or for  $\modalM$) that do not contain subformulas of the form $\lambda:F$.
There is the so-called forgetful translation $^\circ$ from $\formulas^\justificationL$ to $\formulas^\modalL$ given by
\[
\bot^\circ := \bot  \qquad  P^\circ := P  \qquad 
(A \to B)^\circ := A^\circ \to B^\circ  \qquad  ([t]A)^\circ := \Box A^\circ \ .
\]
However, we are mainly interested in the converse direction. A \emph{realization} is a mapping from $\formulas^\modalL$ to $\formulas^\justificationL$ such that for all $A \in \formulas^\modalL$, we have $(r(A))^\circ = A$. 

Now the question is whether a realization theorem holds, i.e.~given a modal logic $\modalL$ and a justification logic $\justificationL$, does there exist a realization $r$ such that for all  $A \in \formulas^\modalL$, we have that $\modalL \vdash A$ implies  $\justificationL \vdash r(A)$\ ?

In order to establish such a realization theorem, we need the notion of a schematic constant specification. 

\begin{definition}
A constant specification $\CS$ is called \emph{schematic} if it satisfies the following: 
for each constant $c$, the set of axioms $\{ A \ |\ (c,A) \in \CS\}$ consists of all instances of one or several (possibly zero) axioms schemes of the justification logic.
\end{definition}

Schematic constant specifications are important in the context of substitutions, where a substitution replaces atomic propositions with formulas, proof variables with proof terms, and justification variables with justification terms. 
The following lemma is standard~\cite{justificationLogic2019}.
\begin{lemma}\label{l:subs:1}
Let $\CS$ be a schematic constant specification. We have for any set of formulas $\Delta$, any formula $A$, and any substitution $\sigma$
\[
\Delta \vdash A 
\quad\text{implies}\quad
\Delta\sigma \vdash A\sigma \ . 
\]
\end{lemma}

In order to show a realization result, we further need a cut-free sequent calculus for the given modal logic. The system $\mathsf{GE}$ is given by the following propositional axioms and rules, the structural rules, and the rule {(\bf RE)}. If we replace {(\bf RE)} with {(\bf RM)}, we obtain the system $\mathsf{GM}$.
In these systems, a \emph{sequent} is an expression of the form $\Gamma \supset \Delta$ where $\Gamma$ and $\Delta$ are finite multisets of formulas.

\par\medskip

\noindent
Propositional axioms and rules: 

\begin{multicols}{2}
\begin{prooftree}
\AxiomC{$P \supset P$}
\end{prooftree}

\begin{prooftree}
\AxiomC{$\Gamma \supset \Delta , A$} 
\AxiomC{$B,\Gamma \supset \Delta$}
\RightLabel{$(\to \supset)$}
\BinaryInfC{$A \to B , \Gamma \supset \Delta$}
\end{prooftree}

\begin{prooftree}
\AxiomC{$\bot \supset $}
\end{prooftree}

\begin{prooftree}
\AxiomC{$A,\Gamma \supset \Delta , B$}
\RightLabel{$(\supset \to)$}
\UnaryInfC{$\Gamma \supset \Delta , A \to B$}
\end{prooftree}

\end{multicols}

\par\medskip
\noindent
Structural rules: 
\begin{multicols}{2}

\begin{prooftree}
\AxiomC{$\Gamma \supset \Delta$}
\RightLabel{$(w \supset)$}
\UnaryInfC{$A, \Gamma \supset \Delta$}
\end{prooftree}

\begin{prooftree}
\AxiomC{$A, A, \Gamma \supset \Delta$}
\RightLabel{$(c \supset)$}
\UnaryInfC{$A, \Gamma \supset \Delta$}
\end{prooftree}

\begin{prooftree}
\AxiomC{$\Gamma \supset \Delta$}
\RightLabel{$(\supset w)$}
\UnaryInfC{$\Gamma \supset \Delta , A$}
\end{prooftree}

\begin{prooftree}
\AxiomC{$\Gamma \supset \Delta , A, A$}
\RightLabel{$(\supset c)$}
\UnaryInfC{$\Gamma \supset \Delta , A$}
\end{prooftree}

\end{multicols}

\par\medskip
\noindent
Modal rules: 

\begin{multicols}{2}

\begin{prooftree}
\AxiomC{$A  \supset B$}
\AxiomC{$B \supset A$}
\RightLabel{({\bf RE})}
\BinaryInfC{$\Box A \supset \Box B$}
\end{prooftree}

\begin{prooftree}
\AxiomC{$A \supset B$}
\RightLabel{({\bf RM})}
\UnaryInfC{$\Box A \supset \Box B$}
\end{prooftree}

\end{multicols}
\par\medskip

The systems $\mathsf{GE}$ and $\mathsf{GM}$ are sound and complete~\cite{Indrzejczak2011,LaLu2000}. 

\begin{theorem}
For each modal logic formula $A$, we have
\begin{enumerate}
\item $\mathsf{GE}  \vdash \ \supset A$ \quad if{f}  \quad$\modalE \vdash A$;
\item $\mathsf{GM}  \vdash \  \supset A$  \quad if{f} \quad $\modalM \vdash A$.
\end{enumerate}
\end{theorem}

\subsection{Realization of the modal logic $\modalE$ in $\logicje$}

To realize the non-normal modal logic $\modalE$ in $\logicje$, we need the following notions.  
Let $\mathcal{D}$ be a  $\mathsf{GE}$-proof of $\supset A$. 
We say that occurrences of $\Box$ in $\mathcal{D}$ are \emph{related} if they occur in the same position in related formulas of premises and conclusions of a rule instance in  $\mathcal{D}$. We close this relationship of related occurrences under transitivity. 

All occurrences of $\Box$ in $\mathcal{D}$ naturally split into disjoint \emph{families} of related $\Box$-occurrences. 

 We call a family of $\Box$-occurrences \emph{essential} if at least one of its members is a $\Box$-occurrence introduced by an instance of ({\bf RE}).  
 
 We say two essential families are \emph{equivalent} if there is an instance of ({\bf RE}) rule which introduces $\Box$-occurrences to each of these two families. 
 
 We close this relationship of equivalent families under transitivity. This equivalence relation makes a partition on the set of all essential families. Hence by a \emph{class of equivalent essential families} we mean the set of all essential families which are equivalent. 

\begin{theorem}[Constructive realization of logic $\modalE$]
For any axiomatically appropriate and schematic constant specification $\CS$, there exist a realization~$r$ such that for each formula $A \in \formulas^\modalL$, we have
 \[
 \mathsf{GE} \vdash \ \supset A \quad \text{implies} \quad \logicje \vdash r(A) \ .
 \]
\end{theorem}
We will not present the full proof of the realization theorem. 
The essence is the same as in the proof of the constructive realization theorem for the Logic of Proofs~\cite{Art01BSL,justificationLogic2019}.

Let $\mathcal{D}$ be the  $\mathsf{GE}$-proof of $\supset A$. 
The realization $r$ is constructed by the following algorithm. We reserve a large enough set of proof variables as \emph {provisional variables}.
 
\begin{enumerate}
\item For each non-essential family of $\Box$-occurrences, replace all occurrences of $\Box$ by $[\e(\xi)]$ such that each family has a distinct proof variable $\xi$. 
\item For a class of equivalent essential families $F$, enumerate all instances of {\bf RE} rules which introduce a $\Box$-occurrence to this class of families. Let $n_f$ denote the number of all such {\bf RE} rule instances. Replace each $\Box$ of this class of families with a justification term $[\e(\zeta_1 +...+ \zeta_{n_f})]$ where each $\zeta_i$ is a provisional variable. 
Applying this step for all classes of equivalent essential families yields a derivation tree $\mathcal{D'}$ labeled by  $\formulas^\justificationL$-formulas.
\item Replace all provisional variables in $\mathcal{D'}$ from the leaves toward the root. By induction on the depth of a node in $\mathcal{D'}$, we show that after each replacement, the resulting sequent of this step is derivable in $\logicje$
where
for finite multisets $\Gamma$ and $\Delta$ of $\formulas^\justificationL$-formulas,  derivability of $\Gamma \supset \Delta$ means $\Gamma \vdash _{\CS} \bigvee \Delta$.

\end{enumerate}

According to the enumeration defined in 2, the $i$th occurrence of {\bf RE} rule in $\mathcal{D'}$ is labelled by:

\begin{prooftree}
\AxiomC{$ A \supset B \ \ \ \ \  B \supset A$}
\RightLabel{$(\bf RE)$}
\UnaryInfC{$ [\e(\kappa_1 +\ldots+ \zeta_i+ \ldots+ \kappa_{n_f})]A \supset [\e(\kappa_1 +\ldots+ \zeta_i+ \ldots+ \kappa_{n_f})]B$}
\end{prooftree}
where the $\kappa$'s are proof terms and $\zeta_i$ is a provisional variable.
\medskip 

By I.H. we have $A \vdash_{\CS} B$ and $B \vdash_{\CS} A$.
 By the Deduction Theorem we get  $\vdash _{\CS} A \to B$ and  $\vdash _{\CS} B \to A$.
  By the internalization lemma there are proof terms such $\lambda_{i_1} , \lambda_{i_2}$ that 
  $\vdash_{\CS} \lambda_{i_1} :(A \to B)$ and $\vdash_{\CS} \lambda_{i_2} :(B \to A)$. 
 Replace $\zeta_i$ globaly in the whole derivation $\mathcal{D'}$ with $(\lambda_{i_1} + \lambda_{i_2} )$.
  
Now by axiom $\jplus$ we conclude
  \[  \vdash_{\CS} (\kappa_1 + \ldots + (\lambda_{i_1} + \lambda_{i_2}) + \ldots + \kappa_{n_f} ) :(A \to B)
  \]
and similarly 
\[ \vdash_{\CS} (\kappa_1 + \ldots + (\lambda_{i_1} + \lambda_{i_2}) + \ldots + \kappa_{n_f} ) :(B \to A).
\]
    By axiom {\bf je} we find  
  \[ 
  \vdash _{\CS} [\e(\kappa_1 + \ldots  + (\lambda_{i_1} + \lambda_{i_2}) + \ldots + \kappa_{n_f})]A \to [\e(\kappa_1 + \ldots  + (\lambda_{i_1} + \lambda_{i_2}) + \ldots + \kappa_{n_f})]B
  \]
Note that since $\CS$ is schematic and by Lemma~\ref{l:subs:1},  replacing $\zeta_i$ with $(\lambda_{i_1} + \lambda_{i_2} )$ in  $\mathcal{D'}$ does not affect already established derivability results.

\subsection{Realization of the modal logic $\modalM$ in $\logicjem$}
In order to realize the modal logic $\modalM$ in $\logicjem$, we need some technical notions about occurrences of $\Box$-operators.

We assign a \emph{positive or negative polarity} to each sub-formula occurrence within a fixed formula $A$ as follows:
\begin{enumerate}
\item To the only occurrence of $A$ in $A$ we assign the positive polarity.
\item If a polarity is assigned to a sub-formula of the form $B \to C$ in $A$, then the same polarity is assigned to $C$ and opposite polarity is assigned to~$B$.
\item If a polarity is already assigned to a sub-formula of the form $\Box B$ in $A$, then the same polarity is assigned to $B$. 
\end{enumerate}

Let $\Box B$ be a sub-formula of  $A$. If $A \in \Delta$ in a sequent $\Gamma \supset \Delta$, then the $\Box$-operator of $\Box B$ has the same \emph{polarity} as the sub-formula occurrence of $\Box B$ in $A$.  If $A \in \Gamma$ in a sequent $\Gamma \supset \Delta$, then the $\Box$-operator of $\Box B$ has the opposite \emph{polarity} as the sub-formula occurrence of $\Box B$ in $A$.

\begin{remark}\label{rem:pol:1}
All rules of $\mathsf{GM}$ respect the polarities of $\Box$-operators. The rule \emph{({\bf RM})} introduces negative $\Box$-occurrence to the left side, and positive $\Box$-occurrence to the right side of the conclusion.  
\end{remark}

In the following we consider the system $\mathsf{GM}$. 
Let $\mathcal{D}$ be  a derivation in $\mathsf{GM}$. 
Again, we say that occurrences of $\Box$ in $\mathcal{D}$ are \emph{related} if they occur in the same position in related formulas of premises and conclusions of a rule instance in~$\mathcal{D}$. We close this relationship of related occurrences under transitivity.

All occurrences of $\Box$ in $\mathcal{D}$ naturally split into disjoint \emph{families} of related $\Box$-occurrences.
We call such a family \emph{essential} if at least one of its members is a positive $\Box$-occurrence introduced by an instance of ({\bf RM}).

Now we are ready to formulate and prove the realization theorem.

\begin{definition}[Normal realization]
A realization is called \emph{normal} if all negative occurrences of\/ $\Box$ are realized by distinct justification variables.
\end{definition}

\begin{theorem}[Constructive realization]\label{th:real:1} 
For any axiomatically appropriate and schematic constant specification $\CS$, there exist a normal realization $r$ such that for each formula $A \in \formulas^\modalL$, we have
 \[
 \mathsf{GM} \vdash \ \supset A \quad \text{implies} \quad \logicjem \vdash r(A) \ .
 \]
\end{theorem}

Let $\mathcal{D}$ be the  $\mathsf{GM}$-proof of $\supset A$. The realization $r$ is constructed by the following algorithm. We reserve a large enough set of justification variables as \emph{provisional variables}.
\begin{enumerate}
\item For each non-essential family of $\Box$-occurrences, replace all occurrences of $\Box$ by $[x]$ such that each family has a distinct justification variable. 
\item For an essential family of $\Box$-occurrences, enumerate all occurrences of ({\bf RM}) rules that introduce a $\Box$-operator to this family. Let $n$ be the number of such occurrences. Replace each $\Box$-occurrence of this family with
$[v_1 +... +v_{n}]$
where each $v_i$ is a fresh provisional variable. Applying this step for all essential families yields a derivation tree $\mathcal{D'}$ labeled by  $\formulas^\justificationL$-formulas.
\item Replace all provisional justification variables in $\mathcal{D'}$ from the leaves toward the root. By induction on the depth of a node in $\mathcal{D'}$, we show that after each replacement, the resulting sequent of this step is derivable in $\logicjem$.

Let us show the case of an instance of ({\bf RM}) with number $i$ in an essential family.
The corresponding node in $\mathcal{D'}$ is labelled by
\begin{prooftree}
\AxiomC{$A \supset B$}
\RightLabel{$({\bf RM})$}
\UnaryInfC{$ [x]A \supset [v_1 + \ldots + v_i + \ldots + v_{n}] B$}
\end{prooftree} 
where the $v$'s are justification terms and $v_i$ is a justification variable.
By I.H.~we get $A  \vdash _{\CS}  B$. By the Deduction Theorem we get $\vdash _{\CS} A \to B$ and Internalization yields a proof term $\lambda$ with $\vdash_{\CS} \lambda :(A \to B)$.
By {\bf jm} we get $\vdash _{\CS} [x]A \to [\m(\lambda , x)]B$. Hence, again by the Deduction Theorem, we find
$[x]A \vdash _{\CS} [\m(\lambda , x)]B$ and thus 
\[
[x]A \vdash _{\CS}   [v_1 + \ldots  + \m(\lambda , x)+ \ldots + v_{n}]B
\]
by axiom $\djplus$.
Substitute $\m(\lambda , x)$ for  $v_i$  everywhere in $\mathcal{D'}$. By Lemma~\ref{l:subs:1} this does not affect the already established derivabilty results since $\CS$ is schematic.
\end{enumerate}

\section{Conclusion}

We have presented two new justification logics $\logicje$ and $\logicjem$ as explicit counterparts of the non-normal modal logics $\modalE$ and $\modalM$, respectively. \\
Having a justification analogue of the modal logic  $\modalE$ is particularly important in deontic contexts since, according to Faroldi~\cite{FaroldiDeonticModals}, deontic modalities are hyperintensional.
Note that $\logicje$ is hyperintensional even if it includes the axiom of equivalence~{\bf je}.  Assume $[\e(\lambda)]F$ and let $G$ be equivalent to $F$.  Then $[\e(\lambda)]G$ only holds if $\lambda$ proves the equivalence of $F$ and $G$.  Thus, in general, for any $\lambda$ with  $[\e(\lambda)]F$ one can find a $G$ such that $G$ is equivalent to $F$ but $[\e(\lambda)]G$ does not hold.

On a technical level, the main novelty in our work is the introduction of two types of terms. This facilitates the formulation of axiom  {\bf je},  which corresponds to the rule of equivalence. Having this principle as an axiom (and not as a rule) in justification logic is important to obtain Internalization (Lemma~\ref{l:internalization:1}).

We have established soundness and completeness of logics $\logicje$ and $\logicjem$ with respect to basic models, modular models and fully explanatory modular models.

We have shown that  for an axiomatically appropriate and schematic constant specification~$\CS$, the justification logics $\logicje$ and $\logicjem$ realize the modal logics $\modalE$ and $\modalM$, respectively.  The realization proof for $\logicjem$ is standard, whereas the realization proof for 
$\logicje$ required some new ideas since the rule ({\bf RE}) does not respect polarities of $\Box$-occurences.

\appendix

\section{Soundness and completeness with respect to basic models}\label{ap:A}

\begin{theorem}[Soundness w.r.t.~basic models] 
The Logic $\logicje$ is sound with respect to basic models.  For an arbitrary constant specification $\CS$ and any formula~$F$, 
\[
\logicje  \prove F \ \Longrightarrow \ \valuation \force F \ \text{for any basic model $\valuation$} \ . 
\] 
\end{theorem}

\begin{proof}
As usual, the proof is by induction on the length of $\logicje$ derivations and a case distinction on the last rule.
The only interesting case is when $F$ is an instance of {\bf je}. Suppose 
\[
\valuation \force \lambda \justifies (A \to B) 
\quad\text{and}\quad
\valuation \force \lambda \justifies (B \to A) 
\quad\text{and}\quad
\valuation \force [\e(\lambda)]A\ .
\] 
Thus we have
\[
(A \to B) \in \valuation(\lambda)
\quad\text{and}\quad
(B \to A) \in \valuation(\lambda)
\quad\text{and}\quad
A \in \valuation(\e (\lambda)) \ .
\] 
By Definition~\ref{d:semOPs:1} we find $B \in \valuation(\lambda) \odot \valuation(\e(\lambda))$. Hence, by the definition of basic model we get
$B \in \valuation(\e(\lambda))$, which is $\valuation \force [\e(\lambda)]B$.
\end{proof}

To prove the completeness theorem, we need to know that $\logicje$ is consistent.

\begin{lemma} 
For any constant specification $\CS$, 
$\logicje$ is consistent.
\end{lemma}
\begin{proof}
As usual, one can show that $\logicje$ is a conservative extension of classical propositional logic. This immediately yields consistency of $\logicje$.
\end{proof}

\begin{definition}
A set of formulas $\Gamma$ is called \emph{$\logicje$-consistent} if for each finite subset $\Sigma \subseteq \Gamma$, we have $\nvdash_{\CS} \bigwedge \Sigma \to \bot$.  The set $\Gamma$  is \emph{maximal $\logicje$-consistent} if $\Gamma$ is consistent and none of its proper supersets is.
\end{definition}

As usual, any consistent set can be extended to a maximal consistent set.
\begin{lemma}[Lindenbaum]
For each $\logicje$-consistent set\/ $\Delta$, there exists a  maximal $\logicje$-consistent set\/ $\Gamma \supseteq \Delta$.
\end{lemma} 

\begin{lemma} 
For any constant specification $\CS$ and maximal $\logicje$-consistent set $\Gamma$, there is a canonical basic model $\varepsilon^c$  induced by $\Gamma$ that is defined as follows:
\begin{description}
\item $\varepsilon^c (P) \colonequals 1, \ \text{if} \ P \in \Gamma \ \text{and} \ \varepsilon^c {(P)} \colonequals 0, \ \text{if} \ P \not\in \Gamma$; 
\item $\varepsilon^c (\lambda) \colonequals \{F  \ | \ \lambda \justifies F \in \Gamma \} $;
\item $\varepsilon^c (t) \colonequals \{F  \ | \ [t]F \in \Gamma \} $.
\end{description}
\end{lemma}

\begin{proof}
First we have to establish that $\varepsilon^c$ is a basic evaluation. We only show the condition
\begin{equation}\label{eq:compProofM:1}
\valuation(\lambda) \odot \valuation(\e(\lambda)) \subseteq \valuation(\e(\lambda)).
\end{equation}
Suppose $B \in \valuation(\lambda) \odot \valuation(\e(\lambda))$, which means there is a formula $A\in \varepsilon^c (\e(\lambda))$ with $(A \to B) \in  \varepsilon^c (\lambda)$ and $(B \to A ) \in \varepsilon^c (\lambda)$.
By the definition of $\varepsilon^c$, we have 
\[
\lambda \justifies (A \to B) \in \Gamma
\quad\text{and}\quad
\lambda \justifies (B \to A) \in \Gamma
\quad\text{and}\quad
[\e(\lambda)]A \in \Gamma \ .
\]
Since $\Gamma$ is a maximal consistent set and 
\[ 
(\lambda \justifies (A \to B) \wedge \lambda \justifies (B \to A) ) \to ( [\e(\lambda)]A \to [\e(\lambda)]B  )
\] 
is an instance of {\bf je}, we obtain $[\e(\lambda)]B \in \Gamma$.
Hence $B \in \varepsilon^c (\e(\lambda))$ and \eqref{eq:compProofM:1} is established.

Next, a truth lemma can be established as usual by induction on formula complexity.  For all formulas $F$,
\begin{equation}\label{eq:TruthLemma:ProofModels}
F \in \Gamma \quad\text{if{f}}\quad \varepsilon^c \Vdash F \ .
\end{equation}

Finally, we show that our basic evaluation $\varepsilon^c$  is factive and hence a basic model. 
Suppose $\varepsilon^c \Vdash \lambda \justifies F$. Hence $ \lambda \justifies F \in \Gamma$. Since $\Gamma$ is maximal consistent, we get by axiom 
{\bf jt} that $F \in \Gamma$. By \eqref{eq:TruthLemma:ProofModels} we conclude   $\varepsilon^c \Vdash F$. 
\end{proof}

Using the Lindenbaum lemma, the canonical basic model and the established truth lemma~\eqref{eq:TruthLemma:ProofModels}, we immediately get the following completeness result.

\begin{theorem}[Completeness w.r.t.~basic models] 
Let $\CS$ be an arbitrary constant specification.
The logic $\logicje$ is complete with respect to basic models. 
For any formula~$F$, 
\[
\logicje \vdash F  \quad\text{if{f}}\quad
\varepsilon \Vdash F  \text{ for all  basic models  $\varepsilon$ for $\logicje$ } \ .
\]
\end{theorem}

\section{Soundness and completeness with respect to modular models}\label{ap:B}

\begin{theorem}[Soundness and completeness w.r.t.~modular models] 
Let $\CS$ be an arbitrary constant specification. 
For each formula $F$ we have
\[
\logicje \prove F \quad\text{if{f}}\quad  \model \force F \text{ for all\/ $\logicje$  modular models $ \model $.}
 \]
\end{theorem}

\begin{proof}
To prove soundness, suppose \mbox{$\model = \langle W, N, \varepsilon \rangle$} is a $\logicje$   modular model, and $\logicje \prove A$. We need to show that $A$ is true in every world $w \in W$. 
Assume that 
$\varepsilon_w$ is a basic model.
Then by soundness with respect to basic models we get  $\varepsilon_w \force A$ and by \eqref{eq:locality:1} we conclude  $\model , w \force  A$.
It remains to show that $\varepsilon_w$ indeed is a basic model, i.e.~that it is factive.
Suppose $\varepsilon_w \force \lambda: F$.
By \eqref{eq:locality:1} we get $\model , w \force \lambda: F$. By factivitiy of modular models we get $\model , w \force F$ and by~\eqref{eq:locality:1} again we conclude  $\varepsilon_w \force F$.

For completeness, suppose that  $\logicje \notprove F$. Since $\logicje$ is complete with respect to basic models, there is a $\logicje$-basic model $\varepsilon$ with $\varepsilon \notforce F$. Now we construct a quasi-model $\model \colonequals \langle \{w\} , N, \varepsilon' \rangle$ with $\varepsilon'_w \colonequals \varepsilon$ and 
\[
N(w) = \{ |G|^\model \ | \ G \in \varepsilon'_w (t) , \ \text {for any} \  t \in \jterms \}.
\] 
By  \eqref{eq:locality:1} we find  $\model , w \notforce  F$. It only remains to show that $\model$ is a modular model:
Factivity follows immediately from~\eqref{eq:locality:1}  and the fact that $\varepsilon$ is factive. 
To show~\eqref{eq:JYB:1}, we
suppose $F \in \varepsilon' _w(t)$.
By the definition of $N$ we get   $|F|^{\model } \in N(w)$, which means $F \in \Box_w$. 
\end{proof}

\section{Soundness and completeness with respect to fully explanatory modular models}\label{ap:C}

The next step is to prove that  $\logicje$  is sound and complete with respect to fully explanatory  $\logicje$ modular models.
Before starting to prove the theorem, we need an auxiliary notion:

\begin{definition}[Proof set] 
Let $\maximal$ be the set of all maximal $\logicje$-consistent sets of formulas. We set
\[
 \maximal := \{ \Gamma  \ | \ \Gamma \ \text{is a maximal} \ \logicje \text{-consistent set} \ \} \ .
 \]
For any formula $F$ we define 
$
\|F\| := \{ \Gamma \ | \ \Gamma \in \maximal \ \text{and} \  F \in \Gamma \} 
$,
called the  \emph{proof set} of $F$. 
\end{definition}

Proof sets share a number of properties, which are given in the following lemma.

\begin{lemma}\label{l:proofSets:1}
For formulas $F , G$ following properties hold:
\begin{enumerate}
\item $\|F \wedge G\| = \|F\| \cap \|G\|$;
\item $\| \neg F\| = \maximal \setminus \|F\|$;
\item $\|F \vee G\| = \|F\| \cup \|G\|$;
\item $ \|F\| \subseteq \|G\| \ \text{iff} \ \vdash F \to G$;
\item $\vdash  (F \leftrightarrow G) \ \text{iff} \ \|F\|=\|G\|$;
\item $\|  \lambda \justifies G \| \subseteq \|G\|$ for any proof term $\lambda$.
\end{enumerate}
\end{lemma}
\begin{proof}
Let only show claim 4.
The claim from right to left immediately follows from closure of maximal consistent sets under modus ponens.
For the other direction,
suppose $ \|F\| \subseteq \|G\|$, but not $\vdash F \to G$. Then $\neg (F \to G)$ is consistent and by Lindenbaum's Lemma there is a maximal consistent set $\Gamma \ni \neg (F \to G)$. This means $F , \neg G \in \Gamma$. Since $F \in \Gamma$ and $ \|F\| \subseteq \|G\|$, we get $G \in \Gamma$, which contradicts $\neg G \in \Gamma$.  
\end{proof}

\begin{theorem}[Soundness and completeness for fully explanatory modular models]
Let $\CS$ be an axiomatically appropriate constant specification. $\logicje$ is sound and complete with respect to  fully explanatory  $\logicje$ modular models.
\end{theorem}

\begin{proof}
Soundness  is a direct consequence of soundness for the class of $\logicje$ modular models. 

To prove completeness, we define a canonical model 
$
 \model ^c \colonequals \langle W^c, N^c, \varepsilon^c \rangle
$
by
\begin{itemize}
\item $W^c \colonequals \maximal$ ;
\item $N^c \colon W^c \to \pset (\pset(W^c))$, such that for each  $\Gamma \in W^c$,
\[
 \|F\| \in N^c (\Gamma)  \text{ iff }  [\e(\gamma)]F \in \Gamma \text{ for some }  \e(\gamma) \in \jterms \ ;
\]
\item $\varepsilon ^c _\Gamma(t)    \colonequals \{F  \ | \ [t]F \in \Gamma \}$ and  $\varepsilon ^c _\Gamma(\lambda)    \colonequals \{F  \ | \ \lambda : F \in \Gamma \}$.
\end{itemize}

Before establishing that this canonical model is a fully explanatory modular model, we show that  the neighborhood function is well-defined. The issue is that different formulas may have the same proof set. Thus we need to show the following lemma.
\begin{lemma}
Let\/ $\CS$ be axiomatically appropriate.
The neighborhood mapping $N^c$ is well-defined: for any $\Gamma \in \maximal$ and any formulas $F , G $, if $\|F\| \in N^c (\Gamma) $ and $\|F\| = \|G\|$,  then there is a term $\e(\lambda) \in \jterms$ such that $[\e(\lambda)]G \in \Gamma$.
\end{lemma}
\begin{proof}
Let $F , G$ be two formulas such that $\|F\| = \|G\|$. For some $\Gamma \in \maximal$, suppose $\|F\| \in N^c( \Gamma)$. By the definition of the canonical model we have
$[\e(\gamma)]F \in \Gamma$  for some   $\e(\gamma) \in \jterms$.
By Lemma~\ref{l:proofSets:1}, we have $\vdash_{\je} F \leftrightarrow G$ and so $\prove_{\je} G \to F $ and $\prove_{\je} F \to G$.  Since $\CS$ is axiomatically appropriate, there are proof terms $\delta_1 , \delta_2$ such that $\vdash_{\je} \delta_1 \justifies (F \to G)$ and $\vdash_{\je} \delta_2 \justifies (G \to F)$. By the {\bf j+} axiom, there is a term $\lambda= (\delta_1 + \delta_2+ \gamma)$ such that $\vdash_{\je} \lambda \justifies (F \to G)$ and $\vdash_{\je} \lambda \justifies (G \to F)$.  By  maximal consistency of $\Gamma$ we get 
\begin{equation}\label{eq:plus:1}
([\e(\lambda)]F \to [\e(\lambda)]G) \in \Gamma
\end{equation}
Further, we get by axiom {\textbf {je+}} and maximal consistency of $\Gamma$ that $[\e(\lambda)]F \in \Gamma$ and thus by \eqref{eq:plus:1} we conclude
$[\e(\lambda)]G \in \Gamma$.
\end{proof}

Next we can establish the truth lemma.
\begin{lemma}[Truth lemma]
For each formula $F$, we have $|F|^{\model^c} = \|F\|$.
\end{lemma}
\begin{proof}
As usual the proof is by induction on  the structure of $F$.
We only show the case when $F$ is $[t]G$. We have the following equivalences:
 $\Gamma \in |[t]G|^{\model^c}$ if{f} 
 $\model^c , \Gamma \force [t]G$  if{f} 
 $G \in \varepsilon^c _\Gamma (t)$  if{f} 
 $[t]G \in \Gamma$  if{f} 
 $\Gamma \in \|[t]G\|$.
\end{proof}

Now we show that the canonical model is a  modular model. 
First, we show that $W^c \not= \emptyset$. Recall that by Lindenbaum's Lemma, for every consistent set of formulas $\Gamma$, there exist a maximally consistent set of formulas that contains $\Gamma$. Since the empty set is consistent, by Lindenbaum's Lemma, there is a maximal consistent set that contains the empty set and is an element of $W^c$. 

Next we show factivity. Suppose $\model^c , \Gamma \force \lambda:G$.
By the truth lemma we get $\lambda:G \in \Gamma$. Since $\Gamma$ is maximally consistent, we obtain by axiom {\bf jt} that $G \in \Gamma$. Again by the truth lemma we conclude $\model^c , \Gamma \force G$.

Now we show that the canonical model satisfies justification yields belief~\eqref{eq:JYB:1}.
Suppose $F \in \varepsilon^c _\Gamma (t)$ for some justification term $t$, some formula~$F$, and some $\Gamma  \in W^c$. 
The term $t$ has the form $\e(\lambda)$ and by the definition of $\varepsilon^c _\Gamma$ we find  $[\e(\lambda)]F \in \Gamma$. 
By the definition of $N^c$ we obtain $\|F\| \in N^c (\Gamma)$. 
Thus, using the the truth lemma, we get  $|F|^{\model ^c} \in N^c (\Gamma)$. 
Thus~\eqref{eq:JYB:1} is established.

It remains to show that the canonical model is fully explanatory.
Suppose $|F|^{\model ^c} \in N^c (\Gamma)$ for some formula $F$ and some $\Gamma   \in W^c$. 
By the truth lemma we find $\|F\| \in N^c (\Gamma)$. 
By the definition of $N^c$, this implies $[t]F \in \Gamma$ for some justification term~$t$.
By the definition of $\varepsilon^c _\Gamma$ we finally conclude $F \in \varepsilon^c _\Gamma (t)$.
\end{proof}

\section{Soundness and completeness with respect to monotonic modular models}\label{ap:Cmon}

\begin{theorem}
 Let $\CS$ be an axiomatically appropriate constant specification. $\logicjem$ is sound and complete with respect to fully explanatory $\logicjem$ monotonic modular models.
\end{theorem}

\begin{proof}
For any set $\mathcal{U} \subseteq \pset(W)$, we say $\mathcal{U}$ is \emph{supplemented} or \emph{monotonic}, if $X \in \mathcal{U} $ and $X \subseteq Y \subseteq W$ then $Y \in \mathcal{U}$. So for any $\mathcal{X} \subseteq \pset(W)$, we denote the closure of $\mathcal{X}$ under supplementation by $(\mathcal{X})^{mon}$. Moreover, a \emph{proof set} is defined as:
 \[
\|F\| := \{ \Gamma \ | \ \Gamma \in \mathsf{M}_{\jem} \ \text{and} \  F \in \Gamma \} 
,\]
where $\mathsf{M}_{\jem}$ is the set of all maximal $\logicjem$-consistent sets.

 Now we define the canonical model $ \model ^c_{mon} \colonequals \langle W^c, N^c _{mon}, \varepsilon^c \rangle$, such that:

\begin{itemize}
 \item $W^c \colonequals \mathsf{M}_{\jem}$ ;
 \item $N^c _{mon} \colonequals (N^c_{min})^{mon}$, such that:
 \[
N^c_{min} (\Gamma) = \{  \|F \| \ | \  [t]F \in \Gamma, \ \text {for some} \ t \in \jterms \} \ ;
 \]
 \item $\varepsilon ^c _\Gamma(t)    \colonequals \{F  \ | \ [t]F \in \Gamma \} $ and  $\varepsilon ^c _\Gamma(\lambda)    \colonequals \{F  \ | \ \lambda : F \in \Gamma \}$.
 \end{itemize}

We will only show that $N^c_{min}$ is well-defined and that  $\model^c_{mon}$ is fully explanatory.  The rest of the completness proof is similar to the case for $\logicje$. 

To establish that $N^c_{min}$ is well defined,  assume that 
$F , G$ are two formulas such that $\|F\| = \|G\|$ with  $\|F\| \in N^c( \Gamma)$ for some  $\Gamma \in\mathsf{M}_{\jem}$.
Thus $[s]F \in \Gamma$ for some justification term $s$.  By Lemma~\ref{l:proofSets:1} we find $\vdash_{\logicjem} F \to G$.
Since $\CS$ is axiomatically appropriate,  there is a proof term $\lambda$ with $\vdash_{\logicjem} \lambda:(F \to G)$.
By axiom  {\bf jm}, we conclude $[\m(\lambda, s)]G \in \Gamma$.

To show that $\model^c_{mon}$ is fully explanatory, suppose $|G|^{\model^c _{mon}}  \in N^c_{mon} (\Gamma)$ for some formula $G$ and some $\Gamma \in \mathsf{M}_{\jem}$.  By truth lemma for $ \model ^c_{mon}$, we have $\|G\| \in  N^c_{mon} (\Gamma)$. By definition of $N^c_{mon}$ it means that either $\|G\| \in N^c_{min} (\Gamma)$ or there exists a formula $H$ such that $\|H\| \in N^c_{min} (\Gamma)$ and $\|H\| \subseteq \|G\|$. In the former case by definition of canonical model $[t]G \in \Gamma$ for some $t \in \jterms$. 
In the latter case,  we find $[t]H \in \Gamma$ for some $t \in \jterms$. 
Moreover,  by Lemma~\ref{l:proofSets:1} we obtain $\prove_{\logicjem} H \to G$. Since $\CS$ is axiomatically appropriate, there is a proof term~$\lambda$ such that $\prove_{\logicjem} \lambda \justifies (H \to G)$. By axiom {\bf {jm}}, there is a term $\m(\lambda, t)$ such that $\prove_{\logicjem} [t]H \to [\m(\lambda, t)]G$. We conclude $[\m(\lambda, t)]G \in \Gamma$. 
\end{proof}

\section{Examples of Realization}\label{ap:D}

\begin{example}
We realize the following theorem of $\modalE$ in $\je$:
 \[ 
 \Box A \to (\Box B \to \Box A).
 \]
Consider the derivation in $\modalE$:
\begin{prooftree}
\AxiomC{$A \supset A$}
\AxiomC{$A \supset A$}
\RightLabel{($\bf {RE}$)}
\BinaryInfC{$\Box A \supset \Box A$}
\RightLabel{$(w \supset)$}
\UnaryInfC{$\Box A , \Box B \supset \Box A$}
\RightLabel{$(\supset \to)$}
\UnaryInfC{$\Box A \supset \Box B \to \Box A$}
\RightLabel{$(\supset \to)$}
\UnaryInfC{$\supset \Box A \to (\Box B \to \Box A)$}
\end{prooftree}
Let $\lambda$ be a proof term such that $\lambda: (A \to A)$ is provable. 
We find the following realization.  Note that ($\bf {je}$) denotes several reasoning steps.
\begin{prooftree}
\AxiomC{$ A \supset A$}
\AxiomC{$ A \supset A$}
\RightLabel{($\bf {je}$)}
\BinaryInfC{$[\e(\lambda)] A \supset [\e(\lambda)] A$}
\RightLabel{$(w \supset)$}
\UnaryInfC{$[\e(\lambda)] A , [\e(\xi_0)] B \supset [\e(\lambda)] A$}
\RightLabel{$(\supset \to)$}
\UnaryInfC{$[\e(\lambda)] A \supset [\e(\xi_0)] B \to [\e(\lambda)] A$}
\RightLabel{$(\supset \to)$}
\UnaryInfC{$\supset [\e(\lambda)]A \to ([\e(\xi_0)] B \to [\e(\lambda)] A)$}
\end{prooftree}
Note that this is already a simplification.  Following the realization procedure exactly as given in the proof, would yield
\[
 [\e(\lambda+\lambda)]A \to ([\e(\xi_0)] B \to [\e(\lambda+\lambda)] A).
 \]
\end{example}

\begin{example}
Realize $\Box \Box A \to \Box \Box A$ in $\je$.
We find the  following  derivation in $\modalE$:
\begin{prooftree}
\AxiomC{$ A \supset A$}
\AxiomC{$ A \supset A$}
\RightLabel{($\bf {RE}$)}
\BinaryInfC{$\Box A \supset \Box A $}
                                                         \AxiomC{$A \supset A$}
                                                         \AxiomC{$A \supset A$}
                                                          \RightLabel{($\bf {RE}$)}
                                                          \BinaryInfC{$\Box A \supset \Box A$}
\RightLabel{($\bf {RE}$)}
 \BinaryInfC{$\Box \Box A \supset \Box \Box A$}
\end{prooftree}
We obtain the following realization, where again $\lambda$ is a proof term with $\lambda: (A \to A)$ and $\kappa$ is a proof term with 
$\kappa: ([\e(\lambda +  \lambda )] A \supset [\e(\lambda +  \lambda)]A )$ being provable.
\begin{prooftree}
\AxiomC{$ A \supset A$}
\AxiomC{$ A \supset A$}
\BinaryInfC{$[\e(\lambda +  \lambda )] A \supset [\e(\lambda +  \lambda )] A$}

                                                         \AxiomC{$ A \supset A$}
                                                         \AxiomC{$ A \supset A$}
                                                            \BinaryInfC{$ [\e(\lambda +  \lambda )] A \supset [\e(\lambda +  \lambda)] A$}
 \BinaryInfC{$ [\e(\kappa)][\e(\lambda +  \lambda )] A \supset
 [\e(\kappa)][\e(\lambda +  \lambda )] A$}
\end{prooftree}
Again, we used a simplification.  The exact procedure would yield 
\[
 [\e(\kappa + \kappa)][\e((\lambda +  \lambda) + (\lambda +  \lambda)  )] A \supset [\e(\kappa + \kappa)][\e((\lambda +  \lambda) + (\lambda +  \lambda) ] A.
 \]
\end{example}

\begin{example}
Realize $\Box (A \to A) \to \Box (B \to B)$ by $\je$. 
We find the  following  derivation in $\modalE$:
\begin{prooftree}
\AxiomC{$B \supset B$}
\RightLabel{$(w \supset )$}
\UnaryInfC{$A \to A , B \supset B$}
\RightLabel{$ (\supset \to)$}
\UnaryInfC{$A \to A \supset B \to B$}
                                                         \AxiomC{$A \supset A$}
                                                         \RightLabel{$(w \supset)$}
                                                             \UnaryInfC{$B \to B , A \supset A$}
                                                              \RightLabel{$ (\supset \to)$}
                                                               \UnaryInfC{$B \to B \supset A \to A$}
\RightLabel{(${\bf RE}$)}
\BinaryInfC{$\Box (A \to A) \supset \Box (B \to B)$}
\RightLabel{$(\supset \to)$}
\UnaryInfC{$\supset \Box (A \to A) \to \Box (B \to B)$}
\end{prooftree}
Let $\lambda_1$  and $\lambda_2$ be proof terms such that $\lambda_1: ((A \to A) \to (B \to B))$ and  $\lambda_2: ((B \to B) \to (A \to A))$ are provable. 
We find the following realization:
\begin{prooftree}
\AxiomC{$B \supset B$}
\RightLabel{$(w \supset )$}
\UnaryInfC{$A \to A , B \supset B$}
\RightLabel{$ (\supset \to)$}
\UnaryInfC{$A \to A \supset B \to B$}
                                                         \AxiomC{$A \supset A$}
                                                         \RightLabel{$(w \supset)$}
                                                             \UnaryInfC{$B \to B , A \supset A$}
                                                              \RightLabel{$ (\supset \to)$}
                                                               \UnaryInfC{$B \to B \supset A \to A$}
\RightLabel{(${\bf je}$)}
\BinaryInfC{$[\e(\lambda_1 + \lambda_2)] (A \to A) \supset [\e(\lambda_1 + \lambda_2)] (B \to B)$}
\RightLabel{$(\supset \to)$}
\UnaryInfC{$\supset [\e(\lambda_1 + \lambda_2)] (A \to A) \to [\e(\lambda_1 + \lambda_2 )] (B \to B)$}
\end{prooftree}

\end{example}

\begin{example}
We realize the axiom scheme {\bf M} : $\Box (A \wedge B) \to (\Box A \wedge \Box B)$ in $\jem$.
We start with its derivation in $\modalM$:
\begin{prooftree}
\AxiomC{$A \supset A$}
\RightLabel{$(w \supset)$}
\UnaryInfC{$A \wedge B \supset A$}
\RightLabel{({\bf RM})}
\UnaryInfC{$\Box(A \wedge B) \supset \Box A$}
                                                                            \AxiomC{$B \supset B$}
                                                                            \RightLabel{$(w \supset)$}
                                                                            \UnaryInfC{$A \wedge B \supset B$}
                                                                            \RightLabel{({\bf RM})}
                                                                           \UnaryInfC{$\Box (A \wedge B) \supset \Box B$}
\RightLabel{$(\to \wedge)$}                                                                        
\BinaryInfC{$\Box (A \wedge B) \supset (\Box A \wedge \Box B)$}  
\RightLabel{$(\supset \to)$}
\UnaryInfC{$\supset \Box (A \wedge B) \to (\Box A \wedge \Box B)$}                                                                          
\end{prooftree}
We find the following realization in $\jem$: 
\begin{prooftree}
\AxiomC{$A \supset A$}
\RightLabel{$(w \supset)$}
\UnaryInfC{$A \wedge B \supset A$}
\RightLabel{$(\supset \to)$}
\UnaryInfC{$A \wedge B \to A$}
\RightLabel{({\bf jm})}
\UnaryInfC{$[x] (A \wedge B) \supset [\m(\lambda , x)] A$}
                                                                                \AxiomC{$B\supset B$}
                                                                                \RightLabel{$(w \supset)$}
                                                                                \UnaryInfC{$A \wedge B \supset B$}
                                                                                \RightLabel{$(\supset \to)$}
                                                                                 \UnaryInfC{$A \wedge B \to B$}
                                                                                \RightLabel{({\bf jm})}
                                                                                \UnaryInfC{$[x] (A \wedge B) \ \supset [\m(\kappa , x)] B$}

\RightLabel{$(\to \wedge)$}
\BinaryInfC{$[x] (A \wedge B) \supset ([\m(\lambda , x)]  A \wedge [\m(\kappa , t)]B)$}
\RightLabel{$(\supset \to)$}
\UnaryInfC{$\supset [x] (A \wedge B) \to ([\m(\lambda , x)]  A \wedge [\m(\kappa , t)]B)$}
\end{prooftree}
where $\lambda , \kappa$ are proof terms with 
\[ \prove_{\jem} \lambda \justifies (A \wedge B \to A) \ \ \text{and} \ \ \prove_{\jem} \kappa \justifies (A \wedge B \to B) \ . 
\]
\end{example}

\begin{example}
Now we consider the formula $\Box A \vee \Box B \to \Box(A \vee B)$ with the following derivation:
\begin{prooftree}
\AxiomC{$A \supset A$}
\RightLabel{$(\supset w)$}
\UnaryInfC{$A \supset A,B$}
\RightLabel{$(\to \vee)$}
\UnaryInfC{$A \supset A \vee B$}
\RightLabel{({\bf RM})}
\UnaryInfC{$\Box A \supset \Box(A \vee B)$}
                                                                        \AxiomC{$B \supset B$}
                                                                        \RightLabel{$(\supset w)$}
                                                                        \UnaryInfC{$B \supset A, B$}
                                                                        \RightLabel{$(\to \vee)$}
                                                                         \UnaryInfC{$B \supset A \vee B$}
                                                                          \RightLabel{({\bf RM})}
                                                                           \UnaryInfC{$\Box B \supset \Box(A \vee B)$}                                                 
\RightLabel{$(\vee \to)$}
\BinaryInfC{$\Box A \vee \Box B \supset \Box (A \vee B)$}                                                                     
\end{prooftree}
We find the following realization tree:
\begin{prooftree}
\AxiomC{$A \supset A$}
\RightLabel{$(\supset w)$}
\UnaryInfC{$A \supset A, B$}
\RightLabel{$(\to \vee)$}
\UnaryInfC{$A \supset A \vee B$}
\RightLabel{$(\supset \to)$}
\UnaryInfC{$\supset A \to A \vee B$}
\RightLabel{$({\bf jm})$}
\UnaryInfC{$[x]A \supset [v_1 + v_2] (A \vee B)$}
                                                                            \AxiomC{$B \supset B$}
                                                                            \RightLabel{$(\supset w)$}
                                                                            \UnaryInfC{$B \supset B, A$}
                                                                            \RightLabel{$(\to \vee)$}
                                                                            \UnaryInfC{$B \supset A \vee B$}
                                                                            \RightLabel{$(\supset \to)$}
                                                                            \UnaryInfC{$\supset B \to A \vee B$}   
                                                                            \RightLabel{$({\bf jm})$}
                                                                            \UnaryInfC{$[y]B \supset [v_1 + v_2] (A \vee B)$}
\RightLabel{$(\vee \to)$}
\BinaryInfC{$[x]A \vee [y]B \supset [v_1 + v_2] (A \vee B)$}
\end{prooftree}
Now we substitute the provisional variables $v_1, v_2$ by terms  $ v_1= \m(\lambda , x)$ and $v_2 = \m(\kappa , y)$  where $\lambda, \kappa$ are proof terms that
\[
\prove_{\jem} \ \lambda \justifies (A \to A \vee B) \ \text{and} \  \prove_{\jem} \ \kappa \justifies (B \to A \vee B) \ .
\]
Hence we obtain
\[ [x]A \vee [y]B \supset [\m(\lambda , x)+ \m(\kappa, y)] (A \vee B) \ . \]
\end{example}

\begin{example}
We realize formula $\Box (\Box A \wedge \Box B) \to (\Box \Box A \wedge \Box \Box B)$ in $\jem$. We start with the following derivation where we do not mention all rule applications.
\begin{prooftree}
\AxiomC{$A \supset A$}
\RightLabel{({\bf RM})}
\UnaryInfC{$\Box A \supset \Box A$}
\RightLabel{$(w \supset)$}
\UnaryInfC{$\Box A \wedge \Box B \supset \Box A$}
\RightLabel{({\bf RM})}
\UnaryInfC{$\Box(\Box A \wedge \Box B) \supset \Box \Box A$}
                                                                            \AxiomC{$B \supset B$}
                                                                            \RightLabel{({\bf RM})}
                                                                            \UnaryInfC{$\Box B \supset \Box B$}
                                                                            \RightLabel{$(w \supset)$}
                                                                            \UnaryInfC{$\Box A \wedge \Box B \supset \Box B$}
                                                                            \RightLabel{({\bf RM})}
                                                                           \UnaryInfC{$\Box (\Box A \wedge \Box B) \supset \Box \Box B$}
\RightLabel{$(\to \wedge)$}                                                                        
\BinaryInfC{$\Box (\Box A \wedge \Box B) \supset \Box \Box A \wedge \Box \Box B$}  
\RightLabel{$(\supset \to)$}
\UnaryInfC{$\supset \Box (\Box A \wedge \Box B) \to (\Box \Box A \wedge \Box \Box B)$}                                                                          
\end{prooftree}
We find the following derivation for suitable proof terms $\lambda_1,  \lambda_2,  \gamma_1,  \gamma_2$:
\begin{prooftree}
\AxiomC{$A \supset A$}
\RightLabel{$({\bf jm})$}
\UnaryInfC{$[x] A \supset [\m(\lambda_1 , x)] A$}
\RightLabel{$(w \supset)$}
\UnaryInfC{$[x] A \wedge [y] B \supset [\m(\lambda_1 , x)] A$}
\RightLabel{$({\bf jm})$}
\UnaryInfC{$[z] ([x] A \wedge [y] B) \supset
$}
\alwaysNoLine
\UnaryInfC{$ 
\qquad\qquad[\m(\gamma_1 , z)] [\m(\lambda_1 , x)] A$}
                                                                            \AxiomC{$B \supset B$}
                                                                            \alwaysSingleLine
                                                                            \RightLabel{$({\bf jm})$}
                                                                            \UnaryInfC{$[y] B \supset [\m(\lambda_2, y)] B$}
                                                                            \RightLabel{$(w \supset)$}
                                                                            \UnaryInfC{$[x]A \wedge [y] B \supset  [\m(\lambda_2 , y)] B$}
                                                                            \RightLabel{$({\bf jm})$}
                                                                           \UnaryInfC{$[z] ([x] A \wedge [y] B) \supset 
                                                                           $}
                                                                           \alwaysNoLine
                                                                           \UnaryInfC{$
                                                                           \qquad\qquad[\m(\gamma_2 , z)]  [\m(\lambda_2 , y)] B$}
                                                                           \alwaysSingleLine
\RightLabel{$(\to \wedge)$}                                                                        
\BinaryInfC{$[z] ([x] A \wedge [y] B) \supset 
([\m(\gamma_1 , z)] [\m(\lambda_1 , x)] A \wedge [\m(\gamma_2 , z)]  [\m(\lambda_2 , y)] B)$} 
\RightLabel{$(\supset \to)$}
\UnaryInfC{$\supset ([z] ([x] A \wedge [y] B) \to
( [\m(\gamma_1 , z)] [\m(\lambda_1 , x)] A \wedge [\m(\gamma_2 , z)]  [\m(\lambda_2 , y)] B))$}                                                                         
\end{prooftree}
\end{example}

\end{document}